\documentclass[12pt]{iopart}
\usepackage{hyperref}
\usepackage{graphicx}
\usepackage{amssymb}

\begin{document}
\title{The collapse of a quantum state as a joint probability construction}
%\author{Name: removed for double-anonymous peer review}
\author{Peter Morgan}
%\address{Address: removed for double-anonymous peer review}
\address{Physics Department, Yale University, New Haven, CT 06520, USA.}
%\ead{Removed for double-anonymous peer review}
\ead{peter.w.morgan@yale.edu}

\begin{abstract}
The collapse of a quantum state can be understood as a mathematical way to construct a joint probability density \emph{even for operators that do not commute}.
We can formalize that construction as a non-commutative, non-associative \emph{collapse product} that is nonlinear in its left operand as a model for joint measurements at time-like separation, in part inspired by the \emph{sequential product} for positive semi-definite operators.
The familiar collapse \emph{picture}, in which a quantum state collapses after each measurement as a way to construct a joint probability density for consecutive measurements, is equivalent to a no-collapse picture in which L\"uders transformers applied to subsequent measurements construct a Quantum-Mechanics--Free-Subsystem of Quantum Non-Demolition operators, not as a dynamical process but as an alternative mathematical model for the same consecutive measurements.
The no-collapse picture is particularly simpler when we apply signal analysis to millions or billions of consecutive measurements.
\end{abstract}
\vspace{5ex}

\noindent Resubmission\scalebox{0.9}[1]{ (with moderate changes) }to\,\emph{J.\,Phys.\,A:\;Math.\,Theor.} on April 10th, 2022.\\
\hspace*{2em}Accepted for publication on May 12th, 2022.\\
\hspace*{2em}Published on July 1st, 2022.\newline
Section numbers, equation numbers, and figure numbers are as in the Published Version, \url{https://doi.org/10.1088/1751-8121/ac6f2f}, but pagination is different.
\maketitle

%\small

\newcommand\Half{{\scriptstyle\frac{\scriptstyle 1}{\raisebox{-0.4ex}{$\scriptstyle 2$}} }}
\newcommand\rmj{{\mathsf{j} }}
\newcommand\Unit{{\hat{\mathbf{1}} }}

\newcommand\CollapseProduct{{\mbox{\normalsize\scalebox{0.6}{\raisebox{0.16em}{$\blacktriangleright$}}\hspace{-0.32em}{$\circ$}}}}
\newcommand\RCollapseProduct{{\mbox{\reflectbox{\normalsize\scalebox{0.6}{\raisebox{0.16em}{$\blacktriangleright$}}\hspace{-0.32em}{$\circ$}}}}}
\renewcommand\CollapseProduct{{{\blacktriangleright}\hspace{-0.35em}{\circ}}}
\renewcommand\RCollapseProduct{{{\circ}\hspace{-0.32em}{\blacktriangleleft}}}
\newcommand\Spectrum[1]{{\mathsf{Spec}[#1]}}
\newcommand\Pz[2]{{\delta(\hat{#1}{-}#2)}}%{{\,\mbox{\raisebox{1.7ex}{$\scriptscriptstyle\bullet$}}\hspace{-1.5ex}{#1}_{#2}}}
\newcommand\PU[2]{{\rme^{\,\rmj#2\hat{#1}}}}
\newcommand\PX[2]{{\,\mbox{\rule[1.65ex]{0.5ex}{0.5ex}}\hspace{-1ex}{#1}_{#2}}}

\newenvironment{myquote}[1]%
  {\list{}{\leftmargin=3em\rightmargin=#1%
           \topsep=0ex\partopsep=0ex\parsep=0ex}\item[]}%
  {\endlist}
\newenvironment{bulletpoints}%
  {\list{$\blacktriangleright$}{\leftmargin=3em%
           \topsep=0ex\partopsep=0ex\parsep=-0.3ex}\item[]}%
  {\endlist}

\section{Introduction}\label{Introduction}
We can model measurement results in multiple ways because of the near duality between states and measurement operators.
Most simply, the Schr\"{o}dinger, Interaction, and Heisenberg ``pictures'' differ by the actions of unitary operators on the state and on measurement operators, but give the same expected measurement results.
We can take the idea of different pictures significantly further by taking collapse of the quantum state, viewed as a L\"uders transformer\cite[\S II.3.1]{BuschGrabowskiLahti}, instead to apply to subsequent measurement operators, so that the state in such a picture does not have to change.
It will be shown here that there are many different such pictures, corresponding to different orderings of nonunitary transformers, and that there are several that are relatively natural, not just one.

\newcommand\CP{{\raisebox{-0.15ex}{${\hspace{-0.16em}\stackrel{\raisebox{-0.5ex}[0ex][-0.5ex]{$\hspace{0.06em}\times$}}{\mathsf{c}}\hspace{-0.1em}}$}}}
\begin{figure}[b]
\centerline{\includegraphics[width=0.9\textwidth]{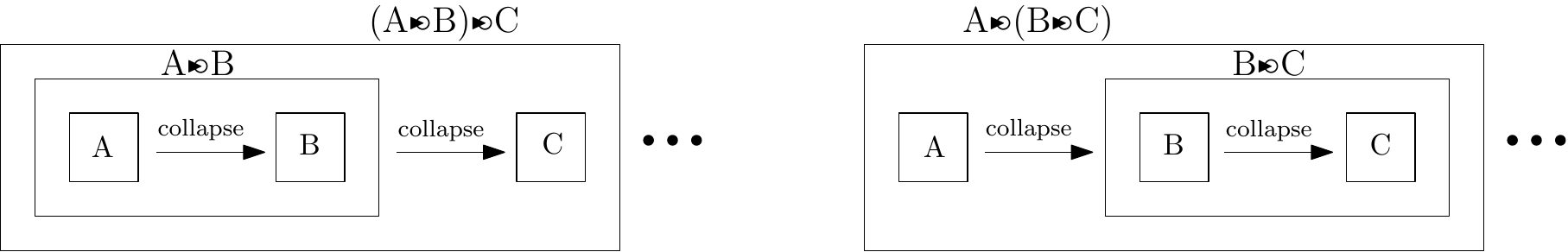}}
\caption{The construction of a joint probability can be understood to use measurements A and B to construct a joint measurement, A-collapse-B (shown in the figure as the noncommutative and nonassociative \emph{collapse product}, A$\CollapseProduct$ B), instead of A collapsing an initial quantum state to a new quantum state that is measured by B.
With another measurement C we can use either (A-collapse-B)-collapse-C or A-collapse-(B-collapse-C), a choice between significantly different alternatives that is forced on us by nonassociativity and that has to be made repeatedly when we consider sequences of many measurements.
That there is such a choice and that it has to be made, explicitly or implicitly, because measurements are very often joint measurements even though they may be modeled by noncommutative operators, has been somewhat hidden by conventions that are associated with different interpretations of quantum mechanics.
Understanding collapse of a quantum state as a way to construct joint probabilities and joint measurements gives us a largely mathematical path to rethinking the measurement problem.\label{JMfigure}}
\end{figure}

It has generally been understood that ``collapse of the wave function'', which we will here call \emph{collapse of the state}, is particularly a feature of quantum mechanics, however as a mathematical tool that allows the construction of joint probability densities even when using noncommutative operators it can be useful in any formalism that uses operators as models for measurement.
The measurement problem can be understood rather differently if we take collapse to be a mathematical construction of joint probability densities and joint measurements instead of as a nonunitary physical process or specific to quantum mechanics (see Fig. \ref{JMfigure}).
As pointed out in ``An algebraic approach to Koopman classical mechanics''\cite{AlgKoopman}, algebraic modeling of measurements and their results, including the use of noncommutativity, can be used in classical mechanics and in signal analysis in the presence of thermal or other noise as effectively as it is in quantum mechanics in the presence of a Poincar\'e invariant quantum noise: for a nontrivial classical dynamics, for example, a Liouville state is in general not an eigenstate both of position measurement and of the Liouvillian operator that generates that classical dynamics.
Even before Koopman's 1931 introduction of a Hilbert space formalism for classical mechanics\cite{Koopman}, Eckart introduced an algebraic approach to classical mechanics in 1926\cite{Eckart}; classical and quantum measurement theory and Wigner function methods can be usefully brought together in such operator and Hilbert space approaches\cite{BondarPRL,BondarPRA,Mukunda}.

For recent reviews of the measurement problem, see, for example, \cite[Ch. 11]{Landsman} and \cite[particularly \S2 and \S\S10-13]{Allahverdyan}.
Both these reviews can be taken to emphasize that \emph{classical measurement theory is incomplete}, as here, almost lamenting that incompleteness,
\begin{quote}{\small ``The right approach, then, must be to define measurement as in the Copenhagen Interpretation, i.e. using a classical description of the apparatus whilst realizing it is ontologically a quantum system''\cite[p. 451]{Landsman}}
\end{quote}
which is echoed by a more historical comment that implies that a quantum apparatus is not classical and cannot be classically described,
\begin{quote}{\small ``In the early days of quantum mechanics, the apparatus was supposed to behave classically, escaping the realm of quantum theory.''\cite[p. 15]{Allahverdyan}}
\end{quote}
The implication that classical measurement theory is incomplete is pervasive in the literature, as here,
\begin{quote}{\small ``As a reaction to the unsolved quantum measurement problem, there is a widespread view that at least part of the macroscopic aspects of an apparatus must be described in terms of classical physics.''\cite[p. 225]{BuschLahtiPellonpaaYlinen}}
\end{quote}
The idea that classical measurement theory should be completed to make it more like quantum measurement theory, by introducing what we might call \emph{Hidden Observables}, is the opposite of a traditional idea that quantum theory should be completed to make it more classical, by introducing \emph{Hidden Variables}, with the problem with the latter being expressed, for example, by Max Jammer,
\begin{quote}{\small ``as long as observation and experiment enforce upon us the present formalism of quantum mechanics, it is logically impossible to complete this formalism to a deterministic description of physical processes''\cite[p. 369]{Jammer}} {\footnotesize(but see also the extensive footnote 16 on that page.)}
\end{quote}
A quantum measurement theory that uses a state and a noncommutative algebra of operators can model the results of experimental procedures and of the algorithms applied to them that cannot be modeled by a traditional classical measurement theory, which uses a state and \emph{only} a commutative algebra of operators.
We can ensure that classical measurement theory is as complete, however, so we do not have to say that a system is ontologically or by description classical or quantum, by using the noncommutative algebra of operators that is provided naturally by the Poisson bracket within Koopman's Hilbert space formalism for classical mechanics.
We will see that a classical noncommutative measurement theory does not have quite as much of a measurement problem because in principle it uses a commutative algebra of operators as a model for joint measurements, however it \emph{can} use noncommutativity when necessary.

We can and should introduce more detailed models of experimental apparatus, for which we can take decoherence\cite{JoosEtAl} or quantum Darwinism\cite{qDarwinism} to be examples of general approaches, however we can take the probability measures that are generated by a quantum state and measurement operators to give us idealized models for statistics, as we do for classical probability, without necessarily knowing anything of the more detailed dynamics, as outlined in \ref{events}.
We particularly do not need to know the detailed thermodynamics of a macroscopic measurement instrument precisely, or of a number of measurement instruments integrated into a whole apparatus, for us to compute relative frequencies from the recorded times of thermodynamic transition events and to construct probability models or ---when measurements are mutually incompatible--- noncommutative probability models for those relative frequencies.

\S\ref{ConventionalCollapse} establishes notation and describes the mathematics of collapse of the quantum state in a form close to what commonly appears in the literature, then \S\ref{CollapseProductSection} proposes an algebraic approach using the \emph{sequential product}\cite{GudderGreechie,Gudder} to construct what will be called here a \emph{collapse product} and \S\ref{EquivalenceSection} exhibits various equivalences of several different collapse+noncommutativity and no-collapse+commutativity approaches to modeling the same joint measurements.
Adopting a no-collapse+commutativity picture is close to the idea of a Quantum-Mechanics--Free-Subsystem introduced by Tsang\&Caves\cite{TsangCaves}, is somewhat prefigured by Belavkin\cite{Belavkin}, and the signal analysis aspect is approached by Anastopoulos\cite{Anastopoulos}, however the use of the sequential product can be understood to put such ideas in a different light.
Adopting a no-collapse+commutativity picture also suggests an alternative approach to the quantum-classical transition\cite{JoosEtAl,qDarwinism,KoflerBrukner,Schlosshauer}, insofar as a system can be completely described within either classical or quantum mechanics ---using whichever is more convenient, as always when we discuss different \emph{pictures}--- \emph{provided} we include the use of noncommutativity and measurement incompatibility in the classical formalism (an extension of the measurement theory of Classical Mechanics that is termed `CM${}_+$' in \cite{AlgKoopman}).
The summary idea suggested here for the various mathematical models is intentionally rather empiricist, taking the actually recorded data as a first foundation, with data and signal analysis leading to many representations of an algebra of measurements together with a quantum state, as described in \S\ref{DSAnalysis}.
Although at two extremes we might say that measurements have underlying hidden causes or that the measurement results are ``spooky'', we can instead commit ourselves to performing new, finer-grained measurements that investigate the subtleties of the circumstances of our previous measurements.
The discussion in \S\ref{MeasurementIncompatibility} points out that in classical mechanics and signal analysis as well as in quantum mechanics some measurement results are not commeasurable, so that a joint probability density that has the required marginal probability densities cannot always be constructed, which was an already known case in the 19th Century, and it is helpful to use noncommutative operators without using the mathematics of collapse to model such cases.

\section{Notation and collapse of a quantum state}\label{ConventionalCollapse}
Linear operators are used in quantum mechanics as models for measurement, with the set of eigenvalues of a self-adjoint operator $\hat A$ corresponding to the sample space of measurement results, together with a \emph{state} $\rho$ that models the average results of past measurements or that fixes what we expect the average results of future measurements will be, which we can write as the expression $\rho(\hat A)$.
For the algebraic conditions satisfied by a state in a quantum mechanical setting, see \ref{TheState} and \cite[\S III.2.2]{Haag}\cite[\S 3.2.1.3]{David}; for accounts of the algebraic approach to quantum mechanics, see \cite[Ch. 3]{David} and \cite{LandsmanAlgQM,Sekhon,Meyer,CliftonBubHalvorson,AlgKoopman}.
When it is possible to introduce a trace map and a density operator, we can write the state as $\rho(\hat A)=\textsf{Tr}[\hat\rho\hat A]$.
Three types of account may be found below: using vector states in an elementary textbook approach, in \ref{JointMeasurement}; using an algebraic formalism, in this section; and using Positive Operator-Valued Measures (POVMs), in \S\ref{POVMContinuousSampleSpace}.

In general, sample spaces associated with actually recorded measurement results for real experiments can always be taken to be discrete sets, because each actual record is always encoded in a finite number of bits.
If we introduce an idealized operator $\hat A$ that has a continuous sample space, we can discretize it using the Heaviside function, so that, as a very coarse-grained example, the operator $\hat A_d=\Theta(\hat A-1)+\Theta(\hat A-2)+\Theta(\hat A-3)$ has the sample space $\{0,1,2,3\}$, corresponding, perhaps, to an actual two bit record.
Much of the literature on the measurement problem restricts itself to the finite sample space case.
\S\ref{POVMContinuousSampleSpace}, however, briefly suggests one approach to working with idealized measurements that have a continuous sample space.

We take a self-adjoint \emph{Moment Generating Operator} $\hat A^\dagger\,{=}\,\hat A$ to have a discrete sample space $\{\alpha_i\}$, so it can be written in a projection-valued presentation as a weighted sum of a complete set of mutually orthogonal projection operators $\hat A=\sum_i\alpha_i\hat P^{(A)}_i$, where $\hat P^{(A)}_i\hat P^{(A)}_j=\delta_{i,j}\hat P^{(A)}_i$ and $\sum_i\hat P^{(A)}_i=\hat{1}$\,\cite[Ch. II]{BuschGrabowskiLahti}\cite{Meyer}, so that $\hat A^n=\sum_i\alpha_i^n\hat P^{(A)}_i$.
From this, we can introduce an abstract complex structure $\rmj$ so that we can construct a 1-parameter group of unitary operators, which we will call here a \emph{Characteristic Function Generating Operator},
\begin{equation}\label{CFGenOp}
  \PU{A}{\lambda}=\sum_i\rme^{\,\rmj\lambda\alpha_i}\hat P_i,\quad\lambda\in\mathbb{R},
\end{equation}
to which we can apply an inverse Fourier transformation to construct a normalized projection-valued distribution, which we will call here a \emph{Probability Density Generating Operator},
\begin{equation}\label{PDGenOp}
  \Pz{A}{u}=\int\rme^{-\rmj\lambda u}\PU{A}{\lambda}\frac{\rmd\lambda}{2\pi}=\sum_i\delta(u-\alpha_i)\hat P^{(A)}_i,\quad u\in\mathbb{R},
\end{equation}
which is a Dirac delta-function form of a Projection-Valued Measure.
As pointed out by Cohen\cite{CohenCF}, characteristic functions are a natural way to work with probability densities in a Hilbert space setting.
In such a setting, one way to introduce the abstract complex structure $\rmj$, as above, pragmatically and following the engineering convention, is as an effective way to manage the ``sine'' and ``cosine'' components of the Fourier transform of a probability density.
The consistent use of $\rmj$ below is intended as something of a reminder that the first place in this algebraic approach where it is natural to introduce an abstract complex structure is specifically associated with characteristic functions, however a complex structure could be introduced for other reasons in a different presentation.

With these constructions, a state generates moments, a probability density, and a characteristic function
\begin{eqnarray}
  &\rho(\hat A^n)&=\,\sum_i\alpha_i^n\rho(\hat P^{(A)}_i),\\
  p(u)=&\;\rho(\Pz{A}{u})\;&=\,\sum_i\delta(u-\alpha_i)\rho(\hat P^{(A)}_i),\quad\mbox{and}\\
  \tilde p(\lambda)=&\rho(\PU{A}{\lambda})&=\,\sum_i\rme^{\rmj\lambda\alpha_i}\rho(\hat P^{(A)}_i),
\end{eqnarray}
for which the probability density is nonzero only for values in the sample space of the measurement.

When two operators commute, which they always do when they are models for measurements that are space-like separated from each other, we can construct a joint probability density straightforwardly, using the ordinary multiplication, as
\begin{equation}
  p(u,v)=\rho(\Pz{A}{u}\,{\cdot}\,\Pz{B}{v})=\sum_{i,j}\delta(u-\alpha_i)\delta(v-\beta_j)\rho(\hat P^{(A)}_i\hat P^{(B)}_j),
\end{equation}
where $\hat B=\sum_j\beta_j\hat P^{(B)}_j\!$,
which extends to any number of commuting operators.
For quantum mechanics, however, operators in general will not commute when they are models for measurements that are time-like separated from each other, in which case $\hat P^{(A)}_i\hat P^{(B)}_j$ may not be a positive operator and we have to use a different construction as a model for the results of sequential measurements, even though they are joint measurements that we can model using joint probability densities.
Following a measurement result $\alpha_i$, we say that the state $\rho$ \emph{``collapses''} to the state
\begin{equation}
  \rho_i(\hat X)=\frac{\rho(\hat P^{(A)}_i\hat X\hat P^{(A)}_i)}{\rho(\hat P^{(A)}_i)},
\end{equation}
giving the expected value for any operator $\hat X$, using the L\"{u}ders operation corresponding to the $i$'th eigenvalue\cite[\S II.3.1]{BuschGrabowskiLahti}.
$\rho_i$ satisfies the four conditions required for it to be a state (as noted above, see \ref{TheState}).
The operation $\rho\mapsto\rho_i$ may also be called a \emph{state reduction} or a \emph{state preparation}.
For a measurement modeled by an operator $\hat B$ the Probability Density Generating Operator $\Pz{B}{v}$ in the state $\rho_i$ then gives a conditional probability density $\rho_i(\Pz{B}{v})$.
The joint measurement probability density after the whole process of a first measurement, collapse of the state, and a second measurement is therefore given by the probability density for the first measurement and the probability densities for the second measurement \emph{conditional} on the result of the first measurement,
\begin{eqnarray}p_{A,\mathrm{collapse},B}(u,v)&=&\sum_i\delta(u-\alpha_i)\rho(\hat P^{(A)}_i)\frac{\rho(\hat P^{(A)}_i\Pz{B}{v}\hat P^{(A)}_i)}{\rho(\hat P^{(A)}_i)}\\
                                                        &=&\sum_{i,j}\delta(u-\alpha_i)\delta(v-\beta_j)\rho(\hat P^{(A)}_i\hat P^{(B)}_j\hat P^{(A)}_i).\label{JointProbAB}
\end{eqnarray}
The sample space $\{(\alpha_i,\beta_j)\}$ of this joint measurement is notably the same as the sample space associated with two commuting operators with sample spaces $\{\alpha_i\}$, $\{\beta_j\}$.
$p_{A,\mathrm{collapse},B}(u,v)$ is positive semi-definite by construction and is normalized, as it must be, $\int\!p_{A,\mathrm{collapse},B}(u,v)\rmd{v}\rmd{u}=1$.
See \ref{JointMeasurement} for an equivalent derivation that follows a more textbook approach.

\section{A \emph{collapse product} of probability density and characteristic function generating operators}\label{CollapseProductSection}
Given the construction above, it is quite natural to use the \emph{sequential product} for positive semi-definite operators that is described by Gudder\&Greechie\cite{GudderGreechie}, $\hat X\circ\hat Y=\sqrt{\hat X}\cdot\hat Y\cdot\!\sqrt{\hat X}$.
For a number of recent articles about the sequential product by Gudder, see \cite{Gudder}.
We can define for the discrete sample space case a \emph{collapse product} of two probability density generating operators or characteristic function generating operators as
\begin{eqnarray}
  \Pz{A}{u}\CollapseProduct\Pz{B}{v}
            %        = \delta(\hat A-u)\CollapseProduct\delta(\hat B-v)
                    &\doteq&\sum_i\delta(u-\alpha_i)\hat P_i^{(A)}\delta(\hat B-v)\hat P_i^{(A)}\cr
                    &=&\sum_{i,j}\delta(u-\alpha_i)\delta(v-\beta_i)\hat P^{(A)}_i\circ\hat P^{(B)}_j,\\
  \PU{A}{\lambda}\CollapseProduct\PU{B}{\mu}
                    &\doteq&\sum_i\rme^{\,\rmj\lambda\alpha_i}\hat P_i^{(A)}\rme^{\,\rmj\mu\hat B}\hat P_i^{(A)}\cr
                    &=&\sum_{i,j}\rme^{\,\rmj(\lambda\alpha_i+\mu\beta_j)}\hat P^{(A)}_i\circ\hat P^{(B)}_j
\end{eqnarray}
(where for projection operators we have $\sqrt{\hat P}=\hat P$), so we can write 
\begin{equation}p_{A,\mathrm{collapse},B}(u,v)=\rho(\Pz{A}{u}\CollapseProduct\Pz{B}{v}),
\end{equation}
and we can define the reverse case, $\Pz{A}{u}\RCollapseProduct\Pz{B}{v}\doteq\Pz{B}{v}\CollapseProduct\Pz{A}{u}$.
We can think of this construction as applying a collapse or L\"{u}ders operation to other measurements instead of to the state, following Bohr's preference for measurements affecting other measurements instead of collapse of the state\cite{Howard}.
We can confirm that $\Pz{A}{u}\CollapseProduct\Pz{B}{v}$ is a positive semi-definite operator and that it is normalized appropriately to generate a joint probability density in any state, $\int \Pz{A}{u}\CollapseProduct\Pz{B}{v}\rmd{u}\rmd{v}=\hat{1}$.
If $[\hat A,\hat B]=0$, then $[\hat P^{(A)}_i,\hat B]=0$, so in that case $\Pz{A}{u}\CollapseProduct\Pz{B}{v}=\Pz{A}{u}\,{\cdot}\,\Pz{B}{v}$.

%Consequently, $\Spectrum{\rme^{\rmj\lambda\hat A}\CollapseProduct\rme^{\,\rmj\mu\hat B}}$ is equal to the spectrum of an operator $\rme^{\rmj(\lambda\hat A+\mu\hat B')}$, where $\Spectrum{\hat B'}=\Spectrum{\hat B}$ and $\hat A$ and $\hat B'$ commute, $[\hat A,\hat B']=0$, $\Spectrum{\PU{A}{\lambda}\CollapseProduct\PU{B'}{\!\!\mu}}=\Spectrum{\PU{A}{\lambda}\PU{B'}{\!\!\mu}}$, and we can define a state $\rho'$ for which $\rho'(\rme^{\rmj(\lambda\hat A+\mu\hat B')})\doteq\rho(\rme^{\rmj\lambda\hat A}\CollapseProduct\rme^{\,\rmj\mu\hat B})$.

We use the sequential product, in particular, so that we can straightforwardly generalize the collapse product (though we should note that the square root of a positive semi-definite operator is only unique if we insist it is positive semi-definite), for example to three or more characteristic function generating operators, as
\begin{equation}(\rme^{\rmj\lambda_1\hat A_1}\CollapseProduct\rme^{\,\rmj\lambda_2\hat A_2})\CollapseProduct\rme^{\,\rmj\lambda_3\hat A_3}
   \doteq\sum_{i,j,k}\rme^{\,\rmj(\lambda_1\alpha^{(1)}_i+\lambda_2\alpha^{(2)}_j+\lambda_3\alpha^{(3)}_k)}(\hat P^{(A_1)}_i\circ\hat P^{(A_2)}_j)\circ\hat P^{(A_3)}_k,
\end{equation}
where we have to include brackets because the collapse product and the sequential product are nonassociative as well as noncommutative.
Furthermore, the collapse product is nonlinear in its left-hand argument and the construction
\begin{equation}
  (\hat X\circ\hat Y)\circ\hat Z\,{=}\,\sqrt{\!\sqrt{\hat X}{\cdot}\hat Y{\cdot}\!\sqrt{\hat X}}\;{\cdot}\;\hat Z\;{\cdot}\sqrt{\!\sqrt{\hat X}{\cdot}\hat Y{\cdot}\!\sqrt{\hat X}}
\end{equation}
is significantly more complicated than
\begin{equation}
  \hat X\circ(\hat Y\circ\hat Z)\,{=}\,\sqrt{\hat X{\cdot}\hat Y}{\cdot}\hat Z{\cdot}\!\sqrt{\hat Y{\cdot}\hat X}.
\end{equation}
We can call the collapse product \emph{power associative} insofar as $\PU{A}{\lambda}\CollapseProduct\PU{A}{\mu}=\PU{A}{\lambda}\PU{A}{\mu}=\PU{A}{(\lambda+\mu)}$ and
\begin{eqnarray}
  \PU{A}{\lambda_1}\CollapseProduct(\PU{A}{\lambda_2}\CollapseProduct\PU{B}{\mu})&=&\PU{A}{\lambda_2}\CollapseProduct(\PU{A}{\lambda_1}\CollapseProduct\PU{B}{\mu})
       =\PU{A}{(\lambda_1+\lambda_2)}\CollapseProduct\PU{B}{\mu}\cr
       &=&(\PU{A}{\lambda_1}\CollapseProduct\PU{A}{\lambda_2})\CollapseProduct\PU{B}{\mu}.
\end{eqnarray}
For probability density generating functions, both $\left(\!\Pz{A_{{}_1}\!}{u_1}\CollapseProduct\Pz{A_{{}_2}\!}{u_2}\right)\!\CollapseProduct\Pz{A_{{}_3}\!}{u_3}$ and $\Pz{A_{{}_1}\!}{u_1}\CollapseProduct\!\left(\!\Pz{A_{{}_2}\!}{u_2}\CollapseProduct\Pz{A_{{}_3}\!}{u_3}\right)$ are positive semi-definite operators and are normalized appropriately to generate a probability density in any state.
Note that $\sqrt{\hat P^{(A)}_i\hat P^{(B)}_j\hat P^{(A)}_i}$ is well-defined because $\hat P^{(A)}_i\hat P^{(B)}_j\hat P^{(A)}_i$ is a positive semi-definite operator, but neither operator is a projection unless it happens that $[\hat P^{(A)}_i,\hat P^{(B)}_j]=0$ for the particular eigenspaces.

Most commonly, the time ordering of measurements is taken to determine the ordering of collapse products, however, as a mathematical construction, the ordering of collapse products is independent of the time ordering of measurement operators, so that we could perfectly well use the collapse product out of time order if we were to find it useful to do so.
Furthermore, time ordering is only a partial ordering for measurements that are associated with overlapping time intervals, in which case time ordering cannot by itself determine the ordering of collapse products, so that the collapse product may introduce some difficult decisions.
Whereas time reversal is straightforward for no-collapse+commutativity models, time reversal for collapse+noncommutativity models in which time-order determines collapse order introduces significant complications.

We can \emph{loosely} consider the collapse product to be a regularized form of the positive semi-definite but \nobreak{unnormalized} construction
\begin{eqnarray}
\hspace*{-3em}\delta(\hat A-u)\delta(\hat B-v)\delta(\hat A-u)&=&\sum_{i,j,k}\delta(u-\alpha_i)\delta(v-\beta_j)\delta(u-\alpha_k)\hat P^{(A)}_i\hat P^{(B)}_j\hat P^{(A)}_k\cr
&\stackrel{\mathcal{N}}{=}&\sum_{i,j,k}\delta(u-\alpha_i)\delta(v-\beta_j)\delta_{i,k}\hat P^{(A)}_i\hat P^{(B)}_j\hat P^{(A)}_k\cr
&=&\sum_{i,j}\delta(u-\alpha_i)\delta(v-\beta_j)\hat P^{(A)}_i\hat P^{(B)}_j\hat P^{(A)}_i\cr
&=&\hspace{0.5em}\Pz{A}{u}\CollapseProduct\Pz{B}{v},
\end{eqnarray}
where the regularization (the equality up to an infinite normalization that is indicated by $\stackrel{\mathcal{N}}{=}$) replaces the improper expression $\delta(u-\alpha_i)\delta(u-\alpha_k)$ by $\delta(u-\alpha_i)\delta_{i,k}$.
We can understand both $\Pz{A}{u}\CollapseProduct$ and $\PU{A}{\lambda}\CollapseProduct$ acting on operators on their right to be a parameterized set of operations in their Kraus representation\cite[\S 7.4]{BuschLahtiPellonpaaYlinen}.

\subsection{POVMs and continuous sample spaces}\label{POVMContinuousSampleSpace}
For a $\sigma$-algebra $\mathcal{F}$, a collection of subsets of a nonempty set $\Omega$ that is closed under the set complement $X\mapsto\Omega{\setminus}X$ and under countable unions and countable intersections, a \emph{Normalized Positive Operator-Valued Measure}\cite[\S II.1.2]{BuschGrabowskiLahti} is an indexed set of operators $\hat E(X)$, with $X\in\mathcal{F}$, for which $\hat E(X)$ is:
\begin{enumerate}
\item Positive semi-definite: $\hat E(X) \ge 0$ for all $X\in\mathcal{F}$;\vspace{-0ex}
\item Normalized: $\hat E(\Omega) = \hat 1$;\vspace{-0ex}
\item a Measure: $\hat E(\cup X_i)=\sum\hat E(X_i)$ for all disjoint sequences $(X_i)\subset\mathcal{F}$.
\end{enumerate}
$\hat E(X)$ is defined so that the action of a state gives a probability measure $\rho(\hat E(X))$, for which (i) $\rho(\hat E(X)) \ge 0$ for all $X\in\mathcal{F}$; (ii) $\rho(\hat E(\Omega)) = 1$; and (iii) $\rho(\hat E(\cup X_i))=\sum\rho(\hat E(X_i))$ for all disjoint sequences $(X_i)\subset\mathcal{F}$.

We can use a Moment Generating Operator $\hat{A}$ and its associated Probability Density Generating Operator $\Pz{A}{u}$ to construct a special case of a POVM, a \emph{Projection-Valued Measure} (PVM), for which all the positive operators used are projection operators, in two straightforward ways, either using the sample space $\Omega=\mathbb{R}$,
$\hat E^{(A)}(X)=\int_X \Pz{A}{u}\rmd u$, or using the sample space $\Omega=\{\alpha_i\}$, $\hat E^{(A)}(X)=\sum_{i:\alpha_i\in X}\hat P^{(A)}_i$.
Using the latter, we can construct a collapse product as
\begin{eqnarray}
\hat E^{(A,\mathrm{collapse},B)}(X\times Y)=\sum_{i:\alpha_i\in X}\sum_{j:\beta_j\in Y}\hat P^{(A)}_i\circ\hat P^{(B)}_j.
\end{eqnarray}
The use of more general POVMs allows the use of smaller Hilbert spaces by differently encoding measurements, which can be practically and conceptually helpful, however Neumark's theorem\cite[\S II.2.4]{BuschGrabowskiLahti} ensures that the abstract algebraic structure of a system of POVMs can always be encoded equally well by PVMs.

As a positive semi-definite operator, $\hat E(X)$ can be written as a sum or integral of positive operators, of the form $\hat Q=\hat A^\dagger\hat A$, so \emph{for a sufficiently complete set} of positive operators $\{\hat Q_\lambda\}$ we can write, \emph{nonuniquely},
\begin{equation}
  \hat E(X) = \sum_\lambda \kappa_\lambda(X)\hat Q_\lambda,
\end{equation}
where $\kappa_\lambda(X)$ must be positive semi-definite and all the $\kappa_\lambda(X)$ and $\hat Q_\lambda$ are nonlinearly constrained by $\hat E(X)$ being normalized and being a measure.
If $\kappa_\lambda(X)=p_\lambda(X)$ is a normalized measure for each $\lambda$, then the constraint $\sum_\lambda\hat Q_\lambda=\hat 1$ is easily shown to be sufficient, and in that case $p(X)=\rho(\hat E(X))=\sum_\lambda p_\lambda(X)\rho(\hat{Q}_\lambda)$ is a convex sum of the probability measures $p_\lambda(X)$, whether the sample space is continuous or discrete.

If we decide to work in an idealized formalism of continuous sample spaces, therefore, instead of using the available instrumental discretization, one way to proceed is to choose a finite set of positive functions $\kappa^{(A)}_\lambda(X)$ ---where $X$ may be an element in a $\sigma$-algebra of subsets of a continuous sample space $\Omega$--- to replace the densities $\delta(u-\alpha_i)$, with each $\kappa^{(A)}_\lambda(X)$ associated with a positive operator $\hat{Q}_\lambda$ that replaces the projection operators $\hat P^{(A)}_i$.
With a sufficiently enlarged set of positive operators $\{\hat Q_\lambda\}$, we can \emph{nonuniquely} extend this construction to any number of POVMs,
\begin{equation}
  \hat E^{(A)}(X)=\sum_\lambda \kappa^{(A)}_\lambda(X)\hat Q_\lambda,\quad\hat E^{(B)}(X)=\sum_\mu \kappa^{(B)}_\mu(X)\hat Q_\mu,\ ...,
\end{equation}
provided we ensure that $E^{(\bullet)}(X)$ is normalized and is a measure, with which we can construct probability measures
\begin{eqnarray}
  p^{(A)}(X)&\,{=}\,\rho(\hat E^{(A)}(X))\,{=}&\,\sum_\lambda \kappa^{(A)}_\lambda(X)\rho(\hat{Q}_\lambda),\cr
  p^{(B)}(X)&\,{=}\,\rho(\hat E^{(B)}(X))\,{=}&\,\sum_\lambda \kappa^{(B)}_\lambda(X)\rho(\hat{Q}_\lambda),\ ...,
\end{eqnarray}
then we can define a collapse product, \emph{relative to the particular choice} $\{\hat Q_\lambda\}$, as
\begin{eqnarray}
  \hat E^{(A,\mathrm{collapse},B)}(X\times Y)&\doteq&\hat E^{(A)}(X)\stackrel{\scriptscriptstyle Q}{\CollapseProduct}\hat E^{(B)}(Y)\cr
     &\doteq&\sum_\lambda \kappa^{(A)}_\lambda(X)\left[\hat{Q}_\lambda\circ\hat E^{(B)}(Y)\right]\cr
     &=&\sum_{\lambda,\mu} \kappa^{(A)}_\lambda(X)\kappa^{(B)}_\mu(Y)\left[\hat{Q}_\lambda\circ\hat Q_\mu\right]\cr
     &=&\sum_{\lambda,\mu} \kappa^{(A)}_\lambda(X)\kappa^{(B)}_\mu(Y)\sqrt{\hat{Q}_\lambda}\hat Q_\mu\sqrt{\hat{Q}_\lambda}.
\end{eqnarray}
We can understand the construction $\hat E^{(A)}(X){\stackrel{\scriptscriptstyle Q}{\CollapseProduct}}$ to be a Completely Positive \emph{Operation}-Valued Measure in its Kraus representation\cite[\S 7.4]{BuschLahtiPellonpaaYlinen}.
We can also think of $\hat E^{(A)}(X){\stackrel{\scriptscriptstyle Q}{\CollapseProduct}}$ as a conditional state preparation, because, for example for $\hat E^{(A)}(X)$, for any state $\rho(\hat M)$, $\rho(\hat E^{(A)}(X){\stackrel{\scriptscriptstyle Q}{\CollapseProduct}}\hat M)/\rho(\hat E^{(A)}(X))$ is also a state, so we could also call ${\stackrel{\scriptscriptstyle Q}{\CollapseProduct}}$ a \emph{preparation product}.

The dependence on a particular choice for $\{\hat Q_\lambda\}$ makes the collapse product less natural for POVMs than it is for the presentation of each PVM in terms of orthogonal projection operators, unless there is additional information about the physics that makes the choice of $\{\hat Q_\lambda\}$ natural as a generating set for a $*$-algebra, with the POVMs as a secondary construction.

\section{Equivalent noncommutative and commutative models}\label{EquivalenceSection}
Because $p_{A,\mathrm{collapse},B}(u,v)$ \emph{is} a joint probability density, with sample space $\{(\alpha_i,\beta_j)\}$, we can certainly introduce operators $\hat A'$ and $\hat B'$ that have the same sample spaces $\{\alpha_i\}$ and $\{\beta_j\}$ as $\hat A$ and $\hat B$ but which commute, $[\hat A',\hat B']=0$, and a different state $\rho_{\mathrm{AB}}$, for which
\begin{equation}
  \rho_{\mathrm{AB}}\bigl(\delta(\hat A'{-}u){\cdot}\delta(\hat B'{-}v)\bigr)\doteq p_{A,\mathrm{collapse},B}(u,v),
\end{equation}
because the joint probability $p_{A,\mathrm{collapse},B}(u,v)$ defines a state over the commutative algebra generated by the self-adjoint operators $\hat A'$ and $\hat B'$.
For a general element $\hat X=\sum_{m,n}\lambda_{m,n}\hat A'{}^m\hat B'{}^n$, $\lambda_{m,n}\in\mathbb{C}$,
\begin{eqnarray}
  \rho_{\mathrm{AB}}(\hat X^\dagger\hat X)&\,{=}\,&\int\!\sum_{m,n}\sum_{m',n'}\lambda_{m,n}^*\lambda_{m',n'}u^{m+m'}v^{n+n'}
               \rho_{\mathrm{AB}}\bigl(\delta(\hat A'{-}u){\cdot}\delta(\hat B'{-}v)\bigr)\rmd u\rmd v\cr
&\,{=}\,&\int\!\sum_{m,n}\sum_{m',n'}\lambda_{m,n}^*\lambda_{m',n'}u^{m+m'}v^{n+n'}\cr
             &&\qquad\times\quad\sum_{i,j}\delta(u-\alpha_i)\delta(v-\beta_j)\rho(\hat P^{(A)}_i\hat P^{(B)}_j\hat P^{(A)}_i)\rmd u\rmd v\cr
&\,{=}\,&\sum_{i,j} (\sum_{m,n}\lambda_{m,n}\alpha_i^m \beta_j^n)^*(\sum_{m',n'}\lambda_{m',n'}\alpha_i^{m'} \beta_j^{n'})\rho(\hat P^{(A)}_i\hat P^{(B)}_j\hat P^{(A)}_i)\cr
&\,{=}\,&\sum_{i,j} |\Lambda_{i,j}|^2\rho(\hat P^{(A)}_i\hat P^{(B)}_j\hat P^{(A)}_i)\ge 0,\cr
&&\hspace{4em}\mbox{where }\Lambda_{i,j}=\sum_{m,n}\lambda_{m,n}\alpha_i^m \beta_j^n.
\label{EquivalentAbelianAlgebra}
\end{eqnarray}
Although we have arrived at this mathematics by a somewhat different motivation, as a way to remodel collapse of the state within the quantum formalism so that it is classically somewhat more natural, the operators $\hat A'$ and $\hat B'$ are then \emph{Quantum Non-Demolition} (QND) measurements relative to each other\cite{TsangCaves,Belavkin}, for which there is a long history and experience in their use.
The account here is algebraically unconcerned about Hilbert space representation, however \ref{JointInteractingMeasurement} presents commutative models of joint measurements in a Hilbert space formalism in a way that is similar to the presentation of collapse of the quantum state in \ref{JointMeasurement}.

\begin{figure}
\includegraphics[width=0.9\textwidth]{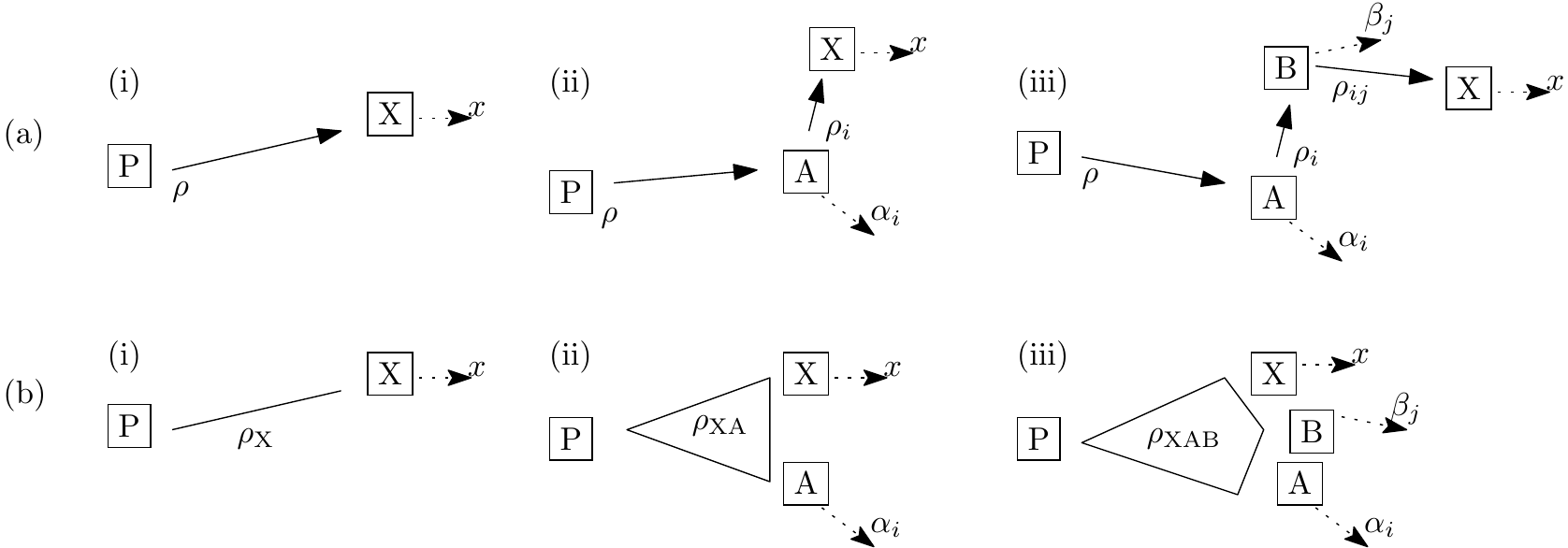}
\caption{\scriptsize(a) \textbf{Collapse}: introducing measurement instruments A and then B does not change the state $\rho$ prepared by the preparation apparatus P \emph{whatsoever}, but the state presented to X changes from $\rho$ to $\rho_i$ and then to $\rho_{ij}$, depending on which measurement results occur for A and B.
Exactly when and how collapse happens to give a joint probability model for the relative frequencies produced in the three cases has been a hard-fought discussion for almost a century.\newline
(b) \textbf{No-collapse}: the state $\rho_X$ is determined by both the preparation apparatus P and the measurement instrument X, which we might attribute to a principle that for every measurement there is a measurement reaction; introducing measurement instruments A and then B changes the boundary conditions of the experimental apparatus, by however much, causing additional reactions, so that the state of the system enclosed by the apparatus changes, $\rho_X\rightarrow\rho_{XA}\rightarrow\rho_{XAB}$, by a corresponding amount, to give a joint probability model for the relative frequencies produced in the three cases.
The operators $\hat X'$, $\hat A'$, and $\hat B'$ that we use as models for the measurements X, A, and B in a no-collapse picture must all mutually commute, so they are different than are used in the collapse picture.\newline
\hspace*{1em}If we think of the apparatus as containing a small number of particles, then it seems unnatural that small changes of the boundary conditions should change the state, but if we \emph{also} think of the apparatus as containing a system of thermodynamically nontrivial waves that is irreducibly noisy at all scales, then it seems natural from a quantum mechanical perspective that every detail of the boundary conditions should contribute to determining the state, just as it would for a thermodynamic state.
The collapse picture's idea that introducing new apparatus does not change the state whatsoever is mathematically practical enough, however, to override any principled qualms about using it.\newline
\hspace*{1em}We can think of X as any environment monitoring or other questions we might ask about P, from nonsense to routine to details at all scales ---``Is there a cat on the optical table?'' ``Is the power on?'' ``Is there dust or any other contaminant on any of the components?''  ``What mechanical and electromagnetic noise spectrum is there in the laboratory?''--- all of which correlate more or less with the results of what we take to be the significant measurements, A and B. We can take X to be part of the state preparation, but that is, so to speak, a different \emph{picture}.\label{PreparationGraphic}}
\end{figure}
Fig. \ref{PreparationGraphic} gives a schematic presentation and discussion of the collapse and no-collapse pictures of a preparation apparatus P, with the environment of the whole experiment measured by X, to which two measurement instruments are added, first A and then B.
The construction in Fig. \ref{PreparationGraphic}(b) is effectively of a quantum state $\rho_{\mathrm{XAB}}$ in what we might call a super-Heisenberg picture, in which the state does not evolve over time and both the collapse and Hamiltonian evolutions are absorbed into the mutually commutative measurement operators that are used as models for the joint measurements X, A, and B. 
%In an algebraic approach, there is no necessity for a Hilbert space representation of the *-algebra of measurement operators, so nothing is said about dimensionality, however if we consider particular representations of the 

\begin{figure}
\includegraphics[width=0.3\textwidth]{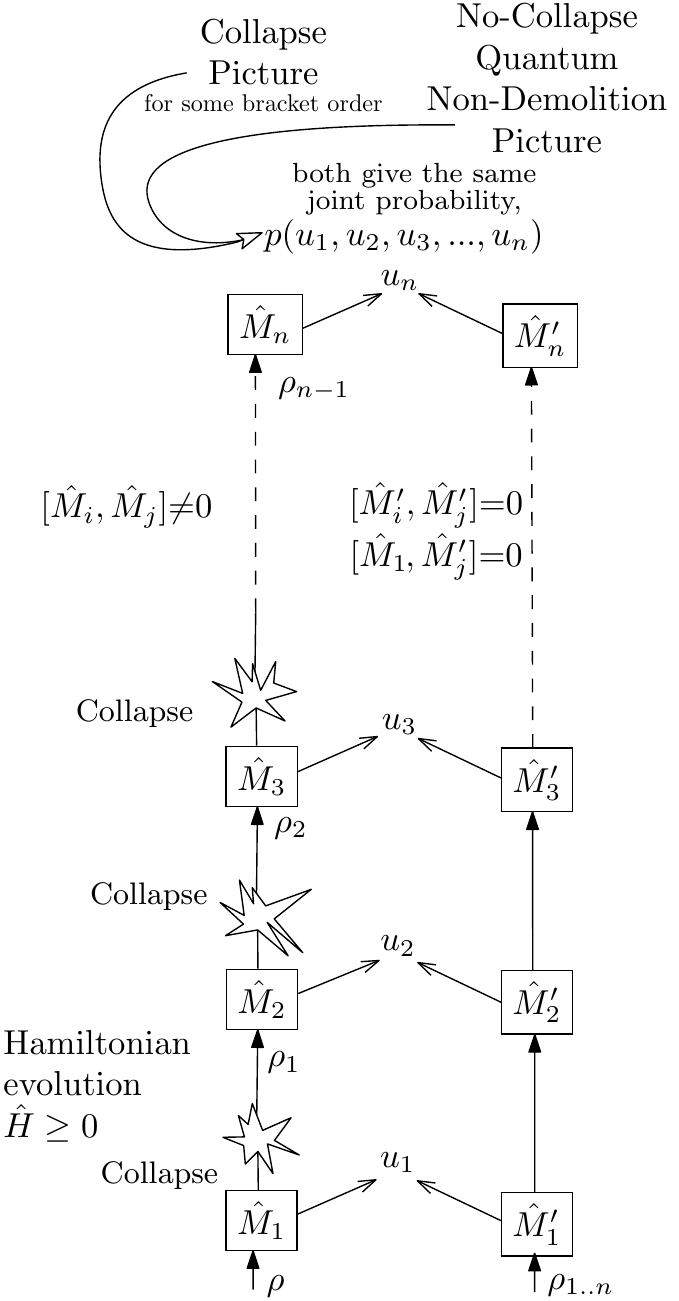}
\caption{\scriptsize When millions or billions of measurements follow one after another, collapse becomes a complicated choice.\label{CollapseGraphic}}
\end{figure}

Particularly when we model a stream of millions or billions of jointly recorded measurement events over time, presented graphically in Fig. \ref{CollapseGraphic}, as we do when we record signal levels on a signal line at regular intervals, it can be equally or more effective simply to use a commutative algebra of measurement operators to model a joint probability density instead of working with a mathematical formalism in which collapse of the state occurs millions or billions of times.
Equally, however, where we have been accustomed to modeling joint probability densities in classical physics only using commutative algebras of measurement operators, it can be justifiable for a classical physicist or in signal analysis to use the collapse product to achieve a useful reduction of Hilbert space dimension or to achieve other goals.

We can also use a single long sequence of noncommutative operators and collapse products in different ways:
\begin{equation}
  (\!{\cdot}{\cdot}{\cdot}((\Pz{A_{{}_1}}{u_1}\CollapseProduct\Pz{A_{{}_2}}{u_2})\CollapseProduct\Pz{A_{{}_3}}{u_3})\CollapseProduct\cdots)\CollapseProduct\Pz{A_{{}_n}}{u_n},
\end{equation}
for example, collapses the state after every measurement, which gives a very different type of model for an experiment than is given by
\begin{equation}
  \Pz{A_{{}_1}}{u_1}\CollapseProduct(\Pz{A_{{}_2}}{u_2}\CollapseProduct(\cdots\CollapseProduct(\delta(\hat{A}_{{}_{n-1}}{-}u_{n-1})
                        \CollapseProduct\Pz{A_{{}_n}}{u_n}){\cdot}{\cdot}{\cdot}\!)),
\end{equation}
which can be thought of as applying a single, combined collapse only after the last-but-one measurement.
We can consider the latter to be the mathematics behind a form of Many Worlds Interpretation, in the sense that there is no collapse except when the experiment is finally completed (which we could say happens when we begin analysis of the actually recorded experimental data, or, more extravagantly, we could say that the final collapse will not happen until after the last human being dies or at some other cosmologically defined endtime.)
Both constructions allow us to generate joint probability densities, because they are both positive semi-definite operators as functions of $u_1, ..., u_n$ and they are both normalized appropriately, but for the same statistics of experimental results we would have to use different states and operators to achieve empirically equivalent models.
A relatively mathematical discussion of the order in which the necessary brackets are or should be applied can equivalently but more implicitly be accommodated by more physically motivated discussions about whether or not collapse happens after every measurement or only after a long sequence of measurements.
Discussions of which parts of the experimental apparatus are considered part of state preparation and which are considered measurement instruments, which can be understood as placing those parts to one side or another of the Heisenberg ``cut'', thereby changing our understanding of what the state describes, can be thought of as a subset of all possible ways in which brackets determine the application of the collapse product.

If we only allow unitary evolution, with no use whatsoever of operations such as collapse of the quantum state, as in the Relative State, Many Worlds, and other interpretations of quantum mechanics, we can only model joint probabilities using commuting operators, giving an essentially classical perspective of QND measurements.
The absolutist idea that only commuting operators should be used for joint probabilities was advocated in \S{7.1} of \cite{AlgKoopman}, with the realization that this is not adequate as a way to understand incompatible measurements and the use of noncommuting operators in quantum mechanics leading to the idea here of the collapse product.
With a no-collapse approach, we are free to adopt any interpretation of classical probability ---Dutch Book, Propensity, Frequency, Many Worlds, or any other--- but there must be a connection to actually recorded experimental data, to the choice of subsets of the data, and to whatever algorithms are applied to construct summary statistics.

Yet another fairly natural construction can be thought of as applying a single, combined collapse in time-reverse order, immediately after the first measurement (taking $\hat A_{{}_i}$ to be before $\hat A_{{}_{i+1}}$),
\begin{equation}
  (\!{\cdot}{\cdot}{\cdot}((\Pz{A_{{}_1}}{u_1}\RCollapseProduct\Pz{A_{{}_2}}{u_2})\RCollapseProduct\Pz{A_{{}_3}}{u_3})
                   \RCollapseProduct\cdots)\RCollapseProduct\Pz{A_{{}_n}}{u_n},
\end{equation}
again requiring a different state and operators to generate the same statistics, but, as already noted, there are many other possibilities.
Indeed, for a given ordering of $n$ distinct operators, there are $\frac{(2n-2)!}{(n-1)!n!}$ distinct ways to introduce the required brackets, giving the sequence $1, 1, 2, 5, 14, 42, 132, ...$, with some choices of brackets being more mathematically natural than others but with none absolutely preferred.

For a given experiment, we can use whichever \emph{picture} seems most helpful, but for each the state will be different and so will our understanding of its relationship to actually recorded experimental data.
The profusion of different states in different pictures, depending on what approach we take to constructing joint probabilities, is straightforward on an epistemological or structural realist understanding of quantum states, however it is arguably incompatible with a na{\"\i}ve ontological understanding of physical states.

\section{Data Analysis and Signal Analysis}\label{DSAnalysis}
A signal analysis approach to classical \emph{or} quantum mechanics takes as its starting point the actually recorded data about the jointly measured signal levels on the many signal lines out of our experimental apparatuses.
For the purposes of such a pragmatic approach, perhaps embedded in a wider philosophy, Megabytes and Terabytes of data is the practical reality, not the signal levels, a vector in a Hilbert space, or anything else, even though such abstractions may be very useful and intuitively helpful.
For such a pragmatic approach, we are relatively little concerned with the wider world that only little affects the recorded data we obtain in a carefully constructed laboratory or observatory.
The process of creating such recorded data is decades or longer in the making: measurement instruments evolve as different materials, materials processing, electronics, and triggers and other algorithms are invented and discarded or adopted.
Some inventions are mostly refinements but others are essentially sidesteps, so that recorded data is never the last word.
Data always has a very long provenance and there is arguably no such thing as \emph{raw} data\cite{BokulichParker,Chang}.
In the recent physics literature, see also \cite{Pronskikh}.

In very broad theoretical terms, we take there to be a number of repeatable measurements $M_1, M_2, M_3, ...$, with the actually recorded data giving us, for each $M_i$, in a na\"{i}ve approach, a sample space $S_i=\{s_{i1}, s_{i2}, ...\}$ and their associated relative frequencies $p_{ij}$.
We can then model those measurements using operators $\hat M_1, \hat M_2, \hat M_3, ...,$ for which the spectrum of each operator is the same as the sample space of the corresponding measurement, $\hat M_i=\sum_j s_{ij}\hat P_{ij}$, for appropriately orthogonal projection operators $\hat P_{ij}$, and we can model the measurement results using a state $\rho$ that gives those relative frequencies, $\rho(\hat P_{ij})=p_{ij}$.
This approach emphasizes how we describe our measurements and relationships between them, \emph{not} how we describe what our measurements are of.

In some cases measurements are clearly \emph{joint} measurements, in which case either we can take the measurement operators to commute, or else we can take them not to commute but we use measurement collapse to construct joint probabilities.
When measurements are \emph{not} joint measurements ---because they correspond, as discussed in \S\ref{MeasurementIncompatibility}, to measurement settings or geometries that cannot be selected in the same time and place--- we should generally not model such measurements using commutative operators, neither classically nor quantum mechanically, even if the corresponding probability densities happen to admit a joint probability, because there may be states that can be prepared for which joint probabilities are not possible.

Less na\"{i}vely, the analysis of actually recorded data, whether in real-time in hardware or software, or in post-analysis by \emph{arbitrary} algorithms, will take into account all our background knowledge of symmetries and other aspects of our experiments as part of decisions about what operators are more appropriate than others as models for particular measurements.
The first level of our knowledge about our experiments can be encoded by imposing additional structure on the index set for measurements: instead of operator models for measurements being indexed by the natural numbers, $\{\hat M_i, i\in\mathbb{N}\}$, we can adopt a more structured index set.
For quantum field theory as laid out by the Wightman axioms, in particular, the index set is typically taken to be a Schwartz space $\mathcal{S}$ of \emph{test functions} on Minkowski space\cite[Ch. II]{Haag}, all of which are smooth and have a smooth Fourier transform, so that we adopt $\{\hat M_f, f\in\mathcal{S}\}$ as an idealized indexed set of measurements.
In signal analysis, the idea of \emph{window functions} in convolution with measurements of the signal level is directly comparable to the idea of test functions.
With symmetries taken as guidelines, the actually recorded data less underdetermines the construction of operators and of a state that models those measurement results.

Where there is space-like separation between measurements, for example, we take it as a well-established empirical principle that operator models for such measurements must commute, to ensure that faster-than-light messaging is not possible, and when measurements are the same except for space-like or time-like translations, rotations, and boosts, we take it as a well-established empirical principle that operator models for such measurements will be related by a unitary representation of the Poincar\'e group.
For the Wightman axioms, these empirical principles have the algebraic consequences that
\begin{enumerate}
\item\label{PoinA}the commutator of operators $\hat M_f$ and $\hat M_g$ must be zero, $[\hat M_f,\hat M_g]\,{=}\,0$, whenever the test functions $f$ and $g$ have space-like separated support;
\item\label{PoinB}where the commutator $[\hat M_f,\hat M_g]$ is non-zero, it must be an operator-valued manifestly Poincar\'e-invariant functional of the test functions $f$ and $g$.
\end{enumerate}
In addition, a Wightman field requires three \emph{a priori} conventions,
\begin{enumerate}\addtocounter{enumi}{2}
\item\label{QFTVacuumState}the \emph{vacuum state} is a Poincar\'e-invariant starting point for the construction of other states, a convention strongly suggested by (\ref{PoinA}) and (\ref{PoinB});
\item\label{PositiveHamiltonian}generators of time-like translations ---Hamiltonian operators--- are required to have a positive spectrum;
\item\label{FieldLinearity}the operator $\hat M_f$ is a linear functional of $f$, $\hat M_{\lambda f+\mu g}\,{=}\,\lambda\hat M_f+\mu\hat M_g$, so that $\hat M_f$ can be constructed in terms of an operator-valued distribution $\hat M(x)$, \makebox{$\hat M_f=\int\hat M(x)f(x)\rmd^4x$.}
\end{enumerate}
If these and other empirical principles and \emph{a priori} conventions are either incorrect or too constrained, however, then there will be some experiments for which it will not be possible to construct adequate models for the actually recorded data and we will have to discover new empirical principles and adopt less restrictive conventions.
The Wightman axioms are well-known to be overconstrained, so that there are no known interacting models in 3+1-dimensions.
Gravity clearly may require modifications to (\ref{PoinA}), (\ref{PoinB}), and (\ref{QFTVacuumState}); the positive spectrum condition, (\ref{PositiveHamiltonian}), is not satisfied for a dynamics associated with QND measurements\cite{TsangCaves} and is not required for the Liouvillian operator that generates time-like translations in classical physics\cite{AlgKoopman}; and the linearity of (\ref{FieldLinearity}) is not an obviously necessary choice for a classical nonlinear theory, because one r\^ole of $\hat M_f$ is to construct modulated states, which casts doubt on its necessity for quantum field theory.

It may seem remarkable that the Wightman axioms have unceremoniously landed here, however it is \emph{axiomatically characteristic} of quantum noise that the quantum vacuum state is Poincar\'e invariant, whereas thermal noise is defined in quantum theory relative to a particular Hamiltonian operator.
Signal analysis, in its simplest aspect of analysis of a single signal level, is a 0+1-dimensional field theory, so no such axiomatic distinction can be made: it is only in a 1+1-dimensional field theory that the mathematical structure of Poincar\'e invariance can be introduced so that we can, in an idealized model for physics, distinguish quantum noise, with an amplitude determined by Planck's constant, from thermal noise, with an amplitude determined by Boltzmann's constant and the temperature.
A distinction commonly made in signal analysis, between the frequency spectra of red, white, and blue noise, requires extension to 1+1-dimensions for a distinction to be made between the wave-number spectra of quantum and thermal noise.
From a classical perspective, the mathematical structure of a measurement theory is rather more that of a generalized classical thermodynamics than of a generalized classical mechanics, partly because the introduction of probability introduces subtleties that are often counterintuitive: the abstract measurements $M_1, M_2, M_3, ...$ could be thought, for example, to be a mathematical model of temperature and density measurements at different places and times.
To my knowledge, there are just three principled generalizations that make a generalized thermodynamics a quantum field theory: (1)~measurement incompatibility [see \S\ref{MeasurementIncompatibility}, immediately below]; (2)~Poincar\'e invariant quantum noise as well as thermal noise; and (3)~the introduction of analyticity by the positive spectrum condition [(\ref{PositiveHamiltonian}) above, but this is universally asserted in axiomatic constructions of quantum mechanics].
A mathematician will be reluctant to give up the superpowers that analyticity affords, however it is notable that the positive spectrum condition is an unnecessary \emph{a priori} convention that significantly modifies discussions of locality.
This brief discussion has introduced a wider mathematical and physical context, which has subtle consequences that cannot be ignored, but the collapse of the quantum state can be largely understood in its narrow mathematical guise of giving one way to construct joint probabilities.

Although we can continue the tradition of taking notable features such as sudden transitions of the signal level on a signal line that are recorded as the times of ``events'' as caused by a ``particle'' or a ``wave'' or a ``quantum particle'' or a ``quantum state'' or a ``quantum field state'', and for the quantum state to ``collapse'' as needed to construct joint probabilities, because we have become skilled in the use of our existing quantum mechanics toolbox, we can also \mbox{---because} of the equivalence of collapse+noncommutativity models and no-collapse+commutativity models constructed \mbox{above---} take the finite collection of actually recorded data about those events to be a consequence of the way the whole experimental apparatus has been engineered.
For future experiments, we can ask a continuum of questions about how statistics of the actually recorded data would differ if small or large changes were made to the apparatus, with arbitrarily many additional measurement instruments and signal lines interpolated among those already there or with the signal levels recorded arbitrarily more accurately or more often.
Loosely, we can think of the recorded past as discrete and finite; we can think of the imagined future, in contrast, as continuous.

\subsection{Measurement Incompatibility}\label{MeasurementIncompatibility}
Joint measurements do not exhaust the use of noncommuting operators in quantum mechanics, because not all ways in which measurements can be combined are joint measurements.
For measurements that are not joint measurements ---because, impossibly, an apparatus setting would have to be different values at the same time or different measurement instruments would have to be in the same place at the same time--- we may have to use noncommuting operators to represent the relationship between different analyses of actually recorded experimental data.
Algebraic formalisms for classical mechanics and signal analysis also can reasonably include the use of noncommuting operators\cite{AlgKoopman}, because the Poisson bracket allows us to generate a noncommutative algebra of transformations that can as reasonably be used as measurement operators in classical mechanics as they are in quantum mechanics.
For experiments in which Bell-CHSH-type inequalities are violated, for example, we \emph{must} include noncommutative operators for an algebraic model to give us an effective model for the results\cite{Landau}\cite[\S 7.2]{AlgKoopman}, because the use of arbitrary post-selection algorithms to create new datasets effectively creates distinct experimental contexts.
To require classical or other operator formalisms not to use noncommutative operators is effectively to make them a straw man.
As Pitowsky puts it\cite[p. 112]{Pitowsky} (saying ``commeasurable'' for ``jointly measurable''), quoted by Abramsky\cite[p. 7]{Abramsky}, 
\begin{quotation}
For certain families of events the theory stipulates that they are commeasurable.
This means that, in every state, the relative frequencies of all these events can be measured on one single sample.
For such families of events, the rules of classical probability ---Boole's conditions in particular--- are valid.
Other families of events are not commeasurable, so their frequencies must be measured in more than one sample.
The events in such families nevertheless exhibit logical relations (given, usually, in terms of algebraic relations among observables).
But for some states, the probabilities assigned to the events violate one or more of Boole's conditions associated with those logical relations.
\end{quotation}
Note that there are other ways of discussing commeasurability: Generalized Probability Theory and other literature often uses the word ``incompatibility'' when two probability densities do not admit a joint probability density\cite{Guhne}, and there is a substantial literature on ``contextuality''\cite{Shahandeh,SchmidEtAl,Ellis}.

\begin{figure}
\includegraphics[width=0.6\textwidth]{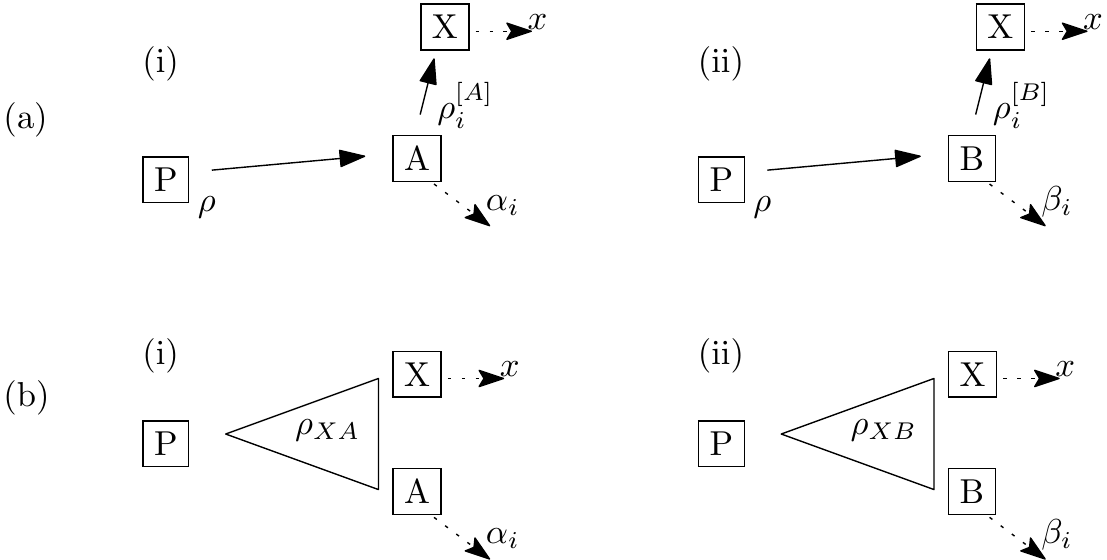}
\caption{\scriptsize(a) \textbf{Collapse}: introducing \emph{either} measurement instrument A \emph{or} B does not change the state $\rho$ prepared by the preparation apparatus P \emph{whatsoever}, but the state presented to X changes either from $\rho$ to $\rho_i^{[A]}$ or to $\rho_i^{[B]}$, depending on which of A or B is introduced.\newline
(b) \textbf{No-collapse}: introducing \emph{either} measurement instrument A \emph{or} B changes the boundary conditions of the experimental apparatus, \emph{differently}, so that the state of the system enclosed by the apparatus changes, $\rho_X\rightarrow\rho_{XA}$ or $\rho_X\rightarrow\rho_{XB}$.\label{IncompatibilityGraphic}}
\end{figure}
Fig. \ref{IncompatibilityGraphic} presents quantum mechanical models for an experimental apparatus that includes a preparation apparatus P and measurement instruments X and either A or B, in a collapse picture and in a no-collapse picture. 
In general, two probability densities,
\begin{eqnarray}
    p_{\mathrm{XA}}(x,u)&=&\rho_{\mathrm{XA}}\bigl(\delta(\hat X{-}x){\cdot}\delta(\hat A{-}u)\bigr)\mbox{\quad and}\cr
    p_{\mathrm{XB}}(x,v)&=&\rho_{\mathrm{XB}}\bigl(\delta(\hat X{-}x){\cdot}\delta(\hat B{-}v)\bigr),
\end{eqnarray}
\emph{cannot} be expected to allow the construction of a joint probability density $p_{\mathrm{XAB}}(x,u,v)$ for which
\begin{eqnarray}
  p_{\mathrm{XA}}(x,u)&=&\int\!p_{\mathrm{XAB}}(x,u,v)\rmd v\mbox{\quad and\quad}\cr
  p_{\mathrm{XB}}(x,v)&=&\int\!p_{\mathrm{XAB}}(x,u,v)\rmd u.
\end{eqnarray}
We \emph{can}, however, construct a single state $\rho_{\mathrm{X\left(A|B\right)}}$ over a non-commutative algebra generated by operators $\hat X$, $\hat A$, and $\hat B$ for which
\begin{eqnarray}
  p_{\mathrm{XA}}(x,u)&=&\rho_{\mathrm{X\left(A|B\right)}}\bigl(\delta(\hat X{-}x){\cdot}\delta(\hat A{-}u)\bigr)\mbox{\quad and\quad}\cr
  p_{\mathrm{XB}}(x,v)&=&\rho_{\mathrm{X\left(A|B\right)}}\bigl(\delta(\hat X{-}x){\cdot}\delta(\hat B{-}v)\bigr),
\end{eqnarray}
exactly as we are familiar with in quantum mechanics, in both Hilbert space and Wigner function formalisms, but this is a mathematical tool that can equally be used when the noise spectrum is different from a quantum noise that has an amplitude that is determined by Planck's constant.
As noted above, a thermal state defines a different from quantum noise spectrum that has an amplitude that is determined by the Boltzmann constant and the temperature.

\section{Discussion}\label{Discussion}
For joint measurements at space-like separation, we do not need to introduce collapse of the state to model them, because operators at space-like separation commute.
Collapse of the state has no effect if measurement operators commute.
For joint measurements at time-like separation, however, the collapse of the state after a measurement makes it \emph{possible} to model those joint measurements using noncommutative operators: without collapse of the state or a similar construction it would not be possible to use noncommutative operators to model joint measurements for all prepared states, already because of the elementary observation that in general $\rho(\hat A\hat B)^*\not=\rho(\hat A\hat B)$ unless $[\hat A,\hat B]=0$.
Any model that uses collapse to make it possible to use noncommutative operators to model joint measurements can be replaced by a model that uses only mutually commuting operators, but, conversely, it also may be useful to replace a commuting operator model for joint probabilities by a model that freely uses a noncommutative algebra of operators.

In the Schr\"odinger picture of phase space quantum mechanics, measurements are associated with regions of space and with a state that only models the statistics of measurement results at one time, evolving unitarily from time to time. In this picture, it is in a sense straightforward to introduce a nonunitary collapse of the quantum state immediately after a measurement result, even though it introduces well-known interpretational concerns.
In contrast, when modeling physical systems in a quantum field theoretic way or in the Heisenberg picture, measurements are associated with regions of space-time and with a state that models the statistics of measurement results at all times, so we cannot as straightforwardly introduce collapse of the quantum state at a particular time, but we can nonetheless use the collapse product acting on other measurements as an algebraic way to construct joint probabilities.

Although it has been stressed here that we \emph{can} think about collapse of the quantum state in terms of the construction of joint probability densities straightforwardly and effectively, the much more elaborate mathematics of detailed models of the thermodynamic behavior and statistical mechanics of real experimental apparatus, of measurement as interaction, of decoherence, and even of the observer's brain or mind, is not thereby made unnecessary.
The many discussions of such mathematics in the literature are of course just as valid and necessary as they ever were, but at the level of abstraction at which we here consider probabilities and relative frequencies of actually recorded experimental results, we can reasonably shut up and calculate joint probabilities using collapse of the state, knowing that we could also shut up and calculate using QND measurements with no collapse if we wished to do so.

The mathematics above allows us to model joint measurements by a commuting algebra of QND measurement operators in any Hilbert space formalism, instead of using collapse of the state or the L\"{u}ders transformer applied to noncommuting operators, but we can use collapse of the state if that gives us interpretative or computational advantages.
As presented in \S\ref{MeasurementIncompatibility}, however, we can and may \emph{have to} use noncommuting operators, in a unification of classical and quantum mechanics, to model measurements that are \emph{not} joint measurements.

\ack
I am grateful for comments from Martin Bohmann, Martin Pl\'avala, David Chester, Denys Bondar, Kadi Harouna Illia, and Spyridon Samios, to Marek Gluza for pressing me on the differences between data analysis and signal analysis, to Richard Gill for reminding me of Belavkin\cite{Belavkin}, to Michael Steiner and Ronald Rendell, whose website and book\cite{SteinerRendell} I have found useful, for a reorienting critique, and to a referee.
The ideas in the paper benefited significantly from presentation as a poster at the W.\&E. Heraeus Workshop on Koopman Methods in Classical and Classical-Quantum Mechanics in April 2021.

\appendix
\section{Idealized logical events}\label{events}
An ``event'' in, for example, an Avalanche PhotoDiode (an APD) can be transformed into idealized logical terms that are very similar indeed to the throw of a coin.
We throw a coin, we see it land, and we record `0' or `1'.
For the APD, we ``throw'' it, a billion times per second, say, then we almost always record a signal level of `0' but very occasionally we record a signal level of `1': this is one aspect of a signal analysis approach that is also elaborated upon in \S\ref{DSAnalysis}.
As for the coin, we have engineered the APD so there is a more-or-less clear coarse-grained distinction between `0' and `1', although there is still, as for the coin, a very small possibility of the `edge' case.
As for the coin, we do not have to solve the problem of how every definite outcome happens to be a `0' instead of a `1', or vice versa, in complete detail, for an idealized formalism of probabilities and relative frequencies to be a worthwhile and enlightening mathematical model.
That we obtain either `0' or `1' as a definite outcome is not something that needs an explanation that ``the state collapsed'', insofar as we engineered the device so that the results would \emph{be} either `0' or `1', just as, for a coin, we do not use a very thick coin that would be more likely to land on its edge, although we can and should, as for the coin, measure in meticulously fine detail how the APD could be constructed or ``thrown'' differently or its surroundings changed to obtain slightly different results.

A more general device will have a less constrained output signal level, however when we record a `1' from the APD there will typically be a few hundred or thousand other `1's either side of it, surrounded by millions of `0's: the data is so sparse that in practice we actually record only the times at which a `1' is first noticed, each as the time of an ``event''.
That compressed, permanently-recorded-on-a-computer classical form of that classical data of `0's and `1', as the times of ``events'', is most of what we have to show to others from an experiment.

The possibility of compression of the results of APD ``throws'', however, as times of ``events'', mathematically differentiates the results of those throws from the results of coin throws.
It introduces the possibility, which can with care be actually realized in experiments, that results of throws in multiple APDs can be \emph{jointly} compressed as the times of ``simultaneous events''.

Further algorithms can be introduced to produce relative frequencies that can be modeled by noncommutative probability theory, which can be as easily accommodated by Koopman's Hilbert space formalism for classical mechanics as it can be by quantum theory's use of Hilbert spaces and operators, so that we do not have to distinguish between classical and quantum systems.
We know very well, however, that the relative frequencies of suitably selected simultaneous events cannot be modeled if we straw-man classical mathematics so that the use of noncommutative probability theory is not allowed\cite{Landau}\cite[\S 7.2]{AlgKoopman}.

\section{Conditions satisfied by a state}\label{TheState}
We take a complex-valued state $\rho$ acting on a $*$-algebra $\mathcal{A}$, $\rho:\mathcal{A}\rightarrow\mathbb{C};\hat A\mapsto\rho(\hat A)$, to satisfy four conditions\cite[\S III.2.2]{Haag}\cite[\S 3.2.1.3]{David}\cite[Ch. 1]{Meyer}:
\begin{itemize}
\item von Neumann complex-linearity: $\rho(\lambda\hat A+\mu\hat B)=\lambda\rho(\hat A)+\mu\rho(\hat B)$, satisfied even if $[\hat A,\hat B]\not=0$;
\item positive semi-definiteness: $\rho(\hat A^\dagger\hat A)\ge 0$;
\item compatibility with the adjoint: $\rho(\hat A^\dagger)=\rho(\hat A)^*$; the adjoint is an antiautomorphism, $(\hat A\hat B)^\dagger=\hat B^\dagger\hat A^\dagger$;
\item normalization: $\rho(\hat 1)=1$.
\end{itemize}
These conditions allow us to use a state to construct a Hilbert space $\mathcal{H}$ and a representation of $\mathcal{A}$ that acts upon it\cite[\S 3]{AlgKoopman}.
Suitable different conditions, which we do not consider here, would allow the construction of Generalized Probability Theories (GPTs) that are not generated by a Hilbert space\cite[Ch. 1]{Holevo}\cite{Janotta,Plavala}.
In an algebraic approach, a state should be distinguished from a ``vector state'', a normalized vector $|\psi\rangle\in\mathcal{H}$, $\langle\psi|\psi\rangle\,{=}\,1$, which can be used to construct a pure state, $\rho_{|\psi\rangle}(\hat A)\,{=}\,\langle\psi|\hat A|\psi\rangle$.
The Born rule expression for a probability density such as
$$|\psi(x)|^2=|\langle x|\psi\rangle|^2\,{=}\,\langle\psi|x\rangle\langle x|\psi\rangle\,{=}\,\rho_{|\psi\rangle}(\hat X)$$
can be loosely understood as the expected measurement result for an operator $\hat X\,{=}\,|x\rangle\langle x|$ in the pure state $\rho_{|\psi\rangle}$.

\section{Joint measurement instruments}\label{JointMeasurement}
We give here a joint measurement instrument account that parallels the more abstract discussion in \S\ref{Introduction}.
Following the account and notation given by Ballentine\cite[\S 3.3]{BallentineFoundPhys} and in \cite[Appendix B]{AlgKoopman}, so that three different levels of discussion are explicitly included here, we consider measurements $\hat A$ and $\hat B$ that have discrete degenerate eigenvalues $a_i$ and $b_j$,
\begin{equation}
    \hat A|a_{i\lambda}\rangle=a_i|a_{i\lambda}\rangle,\quad
    \hat B|b_{j\mu}\rangle=b_j|b_{j\mu}\rangle\label{JMdef}
\end{equation}
(we will omit the degenerate eigenvector indices $\lambda$ and $\mu$ except where necessary.)
To implement these measurements, we introduce measurement instruments A and B that are initially in vector states $|A_0\rangle$ and $|B_0\rangle$ and unitary evolutions
\begin{eqnarray}
  \hat U_{\!{}_A}|a_{i}\rangle\otimes|A_0\rangle&=&|a_{i}\rangle\otimes|A_i\rangle,\label{JMa}\\
  \hat U_{\!{}_B}|b_{j}\rangle\otimes|B_0\rangle&=&|b_{j}\rangle\otimes|B_j\rangle.\label{JMb}
\end{eqnarray}
%We model the measurement instruments using pure states because pure and mixed states are not distinguishable.
By linearity, for a general vector $|\psi\rangle$,
\begin{equation}
  \hat U_{\!{}_A}|\psi\rangle\otimes|A_0\rangle=\sum_i\langle a_i|\psi\rangle\cdot|a_i\rangle\otimes|A_i\rangle,
\end{equation}
and similarly for $\hat U_{\!{}_B}$.
We apply first $\hat U_{\!{}_A}$ and then $\hat U_{\!{}_B}$,
\begin{equation}
  \hat U_{\!{}_B}\hat U_{\!{}_A}|\psi\rangle\,{\otimes}\,|A_0\rangle\,{\otimes}\,|B_0\rangle
    \,{=}\!\sum_j\sum_i\langle b_j|a_i\rangle\langle a_i|\psi\rangle
                                                        \,{\cdot}\,|b_j\rangle\,{\otimes}\,|A_i\rangle\,{\otimes}\,|B_j\rangle,\label{JMab}
\end{equation}
from which, using the Born rule and taking it that $|\psi\rangle$ and the eigenstates are normalized, we extract probabilities
\begin{eqnarray}
  P(A=a_i\,|\psi)&=&|\langle a_i|\psi\rangle|^2\label{JMPa}\\
  P(A=a_i\ \&\ B=b_j\,|\psi)&=&|\langle b_j|a_i\rangle\,\langle a_i|\psi\rangle|^2.\label{JMPab}
\end{eqnarray}
The probability of a measurement result $B=b_j$ given that a measurement $A$ has been made, but averaging over its measurement results, is
\begin{equation}
   P(B=b_j\,|\psi\mbox{ and }A\mbox{ measured})
            =\sum_i|\langle b_j|a_i\rangle\,\langle a_i|\psi\rangle|^2,\label{BwithAmeasured}
\end{equation}
which differs from the probability of a measurement result $B=b_j$ given that a measurement $A$ was never made,
\begin{equation}\label{JMPb}
     P(B=b_j\,|\psi)=\Bigl|\sum_i\langle b_j|a_i\rangle\,\langle a_i|\psi\rangle\Bigr|^2
                             =|\langle b_j|\psi\rangle|^2,
\end{equation}
by the omission of ``interference'' terms, unless $\hat A$ and $\hat B$ commute.
We can rewrite Eq. (\ref{BwithAmeasured}), using a projection operator associated with each eigenvalue $a_i$,
$\hat P^{(A)}_i=\sum_\lambda|a_{i\lambda}\rangle\langle a_{i\lambda}|$, as
\begin{equation}
P(B=b_j\ |\psi\mbox{ and }A\mbox{ measured})
            =\sum_i \langle b_j|\hat P_i^{(A)}|\psi\rangle\,\langle\psi|\hat P_i^{(A)}|b_j\rangle,
\end{equation}
which corresponds to either
\begin{itemize}
\item a L\"uders transformed measurement $\sum_i\hat P_i^{(A)}|b_j\rangle\langle b_j|\hat P^{(A)}_i$ in the state with density matrix $\hat\rho=|\psi\rangle\langle\psi|$, or
\item a measurement $|b_j\rangle\langle b_j|$ in the L\"uders transformed state
$\sum_i\hat P_i^{(A)}|\psi\rangle\langle\psi|\hat P_i^{(A)}$,
\end{itemize}
so, following the algebra, either we can say that a measurement of $B$ after a measurement of $A$ is not in general the same as a measurement of $B$ alone, or we can say that the measurement of $A$ changed the state.
Also introducing $\hat P^{(B)}_j=|b_j\rangle\langle b_j|$, with an implicit sum over degenerate eigenvector indices, we can make an explicit connection to Eq. (\ref{JointProbAB}),
\begin{eqnarray}
  P(A=a_i\ \&\ B=b_j\,|\psi)&=&|\langle b_j|a_i\rangle\,\langle a_i|\psi\rangle|^2\cr
&=&\langle\psi|\hat P^{(A)}_i\hat P^{(B)}_j\hat P^{(A)}_i|\psi\rangle,
\end{eqnarray}
which we can express as a distribution in the state $\rho_{|\psi\rangle}(\hat M)=\langle\psi|\hat M|\psi\rangle$,
\begin{eqnarray}p_{A,\mathrm{collapse},B|\psi}(u,v)&=&\sum_{i,j}\delta(u-a_i)\delta(v-b_j)P(A=a_i\ \&\ B=b_j\,|\psi)\cr
        &=&\sum_{i,j}\delta(u-a_i)\delta(v-b_j)\rho_{|\psi\rangle}(\hat P^{(A)}_i\hat P^{(B)}_j\hat P^{(A)}_i).
\end{eqnarray}

\section{Joint interacting measurement instruments}\label{JointInteractingMeasurement}

If we introduce two measurement instruments A and B, with measurement results $\{a_i\}$ and $\{b_j\}$, we can also say that we have introduced a single measurement instrument AB, with measurement results $\{ab_{ij}\}=\{(a_i,b_j)\}$, whether or not there is any additional consequence whatsoever of the two measurement instruments both being present.
If there is any consequence of A and B both being present, we should at least consider the idea of a measurement instrument AB as something distinct.
Suppose, therefore, in contrast to \ref{JointMeasurement}, that there are three unitary time-like evolutions, $\hat U_{\!{}_{A}}$, $\hat U_{\!{}_{B}}$, and $\hat U_{\!{}_{AB}}$, with no simple relationship between them, instead of writing $\hat U_{\!{}_{AB}}=\hat U_{\!{}_{B}}\hat U_{\!{}_{A}}$.
When both A and B are off (or not present), neither instrument changes state, but when one of A or B is on then there is a set of eigenstates of the prepared system that corresponds to each measurement result for the instrument that is on, $|a_{i\lambda}\rangle\in\mathcal{H}$, $|b_{j\mu}\rangle\in\mathcal{H}$; when both A and B are on then we can in the same way take it that there is a set of eigenstates $|ab_{ij\nu}\rangle\in\mathcal{H}_{\!{}_{AB}}\not=\mathcal{H}$ that corresponds to each joint measurement result and commuting operators $\hat A'$ and $\hat B'$, for which
\begin{eqnarray}
    &&\hat A|a_{i\lambda}\rangle=a_i|a_{i\lambda}\rangle,\hspace{2.9em}
    \hat B|b_{j\mu}\rangle=b_j|b_{j\mu}\rangle,\hspace{9.5em}\mbox{(\ref{JMdef})}\cr
    &&\hat A'|ab_{ij\nu}\rangle=a_i|ab_{ij\nu}\rangle,\quad
    \hat B'|ab_{ij\nu}\rangle=b_j|ab_{ij\nu}\rangle,\cr
    &&\hspace{5em}\hat A'\hat B'|ab_{ij\nu}\rangle=a_ib_j|ab_{ij\nu}\rangle,
\end{eqnarray} 
where the sample space associated with each instrument does not change when both A and B are on, but the Hilbert space $\mathcal{H}_{\!{}_{AB}}$ is in general of higher dimension than the dimension of $\mathcal{H}$ (the degenerate eigenvector indices $\lambda$, $\mu$, and $\nu$ aside, which we will hereafter omit.)
Given that the measurement instruments A, B, and AB that implement these measurements are initially in vector states $|A_0\rangle$, $|B_0\rangle$, and $|A_0\rangle\otimes|B_0\rangle$, and given unitary evolutions
\begin{eqnarray}
\hat U_{\!{}_{A}}|a_i\rangle\hspace{1.3em}\otimes|A_0\rangle\otimes|B_0\rangle &=& |a_i\rangle\hspace{0.75em}\otimes|A_i\rangle\otimes|B_0\rangle,\hspace{6.5em}\mbox{(\ref{JMa})}\cr
\hat U_{\!{}_{B}}|b_j\rangle\hspace{1.3em}\otimes|A_0\rangle\otimes|B_0\rangle &=& |b_j\rangle\hspace{0.75em}\otimes|A_0\rangle\otimes|B_j\rangle,\hspace{6.5em}\mbox{(\ref{JMb})}\cr
\hat U_{\!{}_{AB}}|ab_{ij}\rangle\otimes|A_0\rangle\otimes|B_0\rangle &=& |ab_{ij}\rangle\otimes|A_i\rangle\otimes|B_j\rangle,
\end{eqnarray}
then for a general vector $|\psi_{AB}\rangle\in\mathcal{H}_{\!{}_{AB}}$ we have
\begin{equation}
  \hat U_{\!{}_{AB}}|\psi_{AB}\rangle\,{\otimes}\,|A_0\rangle\,{\otimes}\,|B_0\rangle
    \,{=}\!\sum_j\sum_i\langle ab_{ij}|\psi_{AB}\rangle\cdot|ab_{ij}\rangle\,{\otimes}\,|A_i\rangle\,{\otimes}\,|B_j\rangle.
\end{equation}
In contrast, for $\hat U_{\!{}_{B}}\hat U_{\!{}_{A}}$ and $|\psi\rangle\in\mathcal{H}$, we have
$$  \hat U_{\!{}_B}\hat U_{\!{}_A}|\psi\rangle\,{\otimes}\,|A_0\rangle\,{\otimes}\,|B_0\rangle
      =\sum_j\sum_i\langle b_j|a_i\rangle\langle a_i|\psi\rangle
                                                        \,{\cdot}\,|b_j\rangle\,{\otimes}\,|A_i\rangle\,{\otimes}\,|B_j\rangle.\hspace{2em}\mbox{(\ref{JMab})}$$
Therefore, using the Born rule as in \ref{JointMeasurement}, we have the probabilities
\begin{eqnarray}
P(A\,{=}\,a_i |\psi, \mbox{\small A is on and B is off}) &\,{=}\,& |\langle a_i|\psi\rangle|^2,\strut\hspace{3em}\mbox{(\ref{JMPa})}\cr
P(B\,{=}\,b_j |\psi, \mbox{\small A is off and B is on}) &\,{=}\,& |\langle b_j|\psi\rangle|^2,\strut\hspace{3em}\mbox{(\ref{JMPb})}\cr
P(A\,{=}\,a_i\,{\&}\,B\,{=}\,b_j |\psi_{\!{}_{AB}}, \mbox{\small A is on and B is on}) &\,{=}\,& |\langle ab_{ij}|\psi_{\!{}_{AB}}\rangle|^2,\strut\\
\hspace*{-2.5em}
P(A\,{=}\,a_i\,{\&}\,B\,{=}\,b_j\,|\psi, \mbox{\small A is on,{\footnotesize\,collapse,\,}and B is on})&\,{=}\,&|\langle b_j|a_i\rangle\,\langle a_i|\psi\rangle|^2.\hspace{0.3em}\mbox{(\ref{JMPab})}\nonumber
\end{eqnarray}
The eigenstates $|ab_{ij}\rangle$ are mutually orthonormal, so we can certainly find vectors $|\psi_{\!{}_{AB}}\rangle$ such that $|\langle ab_{ij}|\psi_{\!{}_{AB}}\rangle|^2=|\langle b_j|a_i\rangle\,\langle a_i|\psi\rangle|^2$, for any vector $|\psi\rangle$, effectively as a Hilbert space presentation of the abstract algebraic proof given by Eq. (\ref{EquivalentAbelianAlgebra}).
This is only a minimal constraint on $|\psi_{\!{}_{AB}}\rangle$, which can be more constrained by performing measurements that are incompatible with the measurement instrument AB.

As emphasized in \S\ref{MeasurementIncompatibility}, in general there may not exist a joint probability conditionalized on {\small``only one of A and B is on''} for which the marginals are
\begin{eqnarray*}
  \hspace*{-4em}P(A\,{=}\,a_i |\psi, \mbox{\small A is on and B is off}) &\,{=}\,& \sum_j P(A\,{=}\,a_i\,{\&}\,B\,{=}\,b_j |\mbox{\small only one of A and B is on}),\cr
  \hspace*{-4em}P(B\,{=}\,b_j |\psi, \mbox{\small A is off and B is on}) &\,{=}\,& \sum_i P(A\,{=}\,a_i\,{\&}\,B\,{=}\,b_j |\mbox{\small only one of A and B is on}),
\end{eqnarray*}
if the conditionalizations on ($\psi$, {\small``A is on and B is off''}), on ($\psi$, {\small``A is off and B is on''}), and on {\small``only one of A and B is on''} are sufficiently independent.

The joint evolution $\hat U_{\!{}_{AB}}$ models the relationship between the state preparation and the joint measurement instrument AB as a joint process, instead of taking the measurement instrument A to be prior to the measurement instrument B.
Even if a particle-inspired idea about what happens inside the experimental apparatus thinks of a particle being registered first by the measurement instrument A and being registered second by the measurement instrument B, we can nonetheless \emph{also} consider an alternative model that is less causal, less particle-inspired, and more algebraic, in which there is a joint registration of measurement results in the joint measurement instrument AB.
This and other models may be helpful to have available for those times when particle-inspired ideas about what happens inside an experimental apparatus lead to confusion.

\end{document}